\documentclass[conference]{IEEEtran}
\IEEEoverridecommandlockouts
\usepackage{cite}
\usepackage{amsmath,amssymb,amsfonts}
\usepackage{algorithmic}
\usepackage{graphicx}
\usepackage{textcomp}
\usepackage{xcolor}
\usepackage{subfigure}
\def\BibTeX{{\rm B\kern-.05em{\sc i\kern-.025em b}\kern-.08em
    T\kern-.1667em\lower.7ex\hbox{E}\kern-.125emX}}
\begin{document}
\graphicspath{{./figure/}}

\makeatletter
\newcommand{\rmnum}[1]{\romannumeral #1}
\newcommand{\Rmnum}[1]{\expandafter\@slowromancap\romannumeral #1@}
\makeatother

\title{Fine-grained Modulation for Zigbee Codeword Translation\\

\thanks{Identify applicable funding agency here. If none, delete this.}
}

\author{\IEEEauthorblockN{Kailai Yan}
\IEEEauthorblockA{University of Science and Technology of China \\
yankailai@mail.ustc.edu.cn}
}

\maketitle

\begin{abstract}
In Zigbee backscatter systems, tags piggyback information by adding phase shifts to the RF carriers. The instantaneous-phase shift (IPS) modulation adds phase shifts by toggling between discrete phases, which is easy to realize and is widely used in previous systems. However, as the spectrum efficiency of IPS is poor, it is not suitable for large networks. Thus, frequency-phase shift (FPS) modulation was proposed to make up this drawback. It adds continuous phase shifts to the non-productive carriers by toggling between square waves of different frequencies and has higher spectrum efficiency. In this paper, we realized IPS modulation and FPS modulation on Zigbee single-tone signals, respectively. In addition, we creatively proposed to apply FPS modulation to codeword translation.

We prototype the system using a microchip transmitter, an off-the-shelf FPGA, and a commodity
Zigbee receiver. Through extensive experiments, we proved that FPS modulated Zigbee transmissions have a bandwidth of 1.8 MHz, Which is 2x lower than that of IPS. In addtion, we conducted simulation experiments in MATLAB and demonstrated that FPS modulation can be used in Zigbee codeword translation. 

\end{abstract}


\section{Introduction}
The Internet of Things(IoT) is considered to be a promising research direction and have attracted more and more researchers' attention. However, as the active IoT radios need to produce radio-frequency (RF) carrier by themselves, the energy consumption is too high, which limits the widespread deployment of active IoT radios.

These years, backscatter communication has become popular for its ultra-low power consumption. Instead of actively generating RF signals, the tags modify existing ambient signals to embed tag data, which significantly reduces the power consumption. Specifically, by toggling the reflection parameter, the backscatter tags can produce baseband signals. Then, it mixes the baseband signals with the carriers to modify the information (e.g., amplitude, phase, frequency) of carriers and embed tag data. At last, it transmits the modified carriers to the receiver.

The instantaneous-phase shift (IPS) modulation is a popular modulation scheme to add phase shifts to carriers and is widely used in advanced backscatter systems, such as interscatter. By toggling between square waves of different discrete phases(e.g.,0,+$\frac{\pi}{2}$,+$\pi$,+$\frac{3\pi}{2}$), IPS can add phase shifts in instant. However, as IPS modulation can generate serious sidelobes and has low spectrum efficiency, it is not suitable for large internets. 
Thus, to make up this drawback, frequency-phase shift (FPS) modulation was proposed. The core idea of FPS modulation is to add phase shift to carriers by modifying the frequency. To be more specific, if we want to add a continuous phase shift of $\Delta\phi$ within the transition time \textit{$\Delta$T}, we should shift the carrier frequency by \textit{f ( f = $\frac{\Delta \phi }{2\pi \Delta T}$)}, and keep it for a time of \textit{$\Delta$T}. The tag generates square waves with frequency f and mixes them with carriers to achieve the frequency shift. As FPS modulation is a kind of fine-grained phase modulation, it can be used for many wireless protocols (e.g., bluetooth, wifi, zigbee).

\begin{figure}[tbp]
\centering
\centerline{\includegraphics[width=0.95\linewidth]{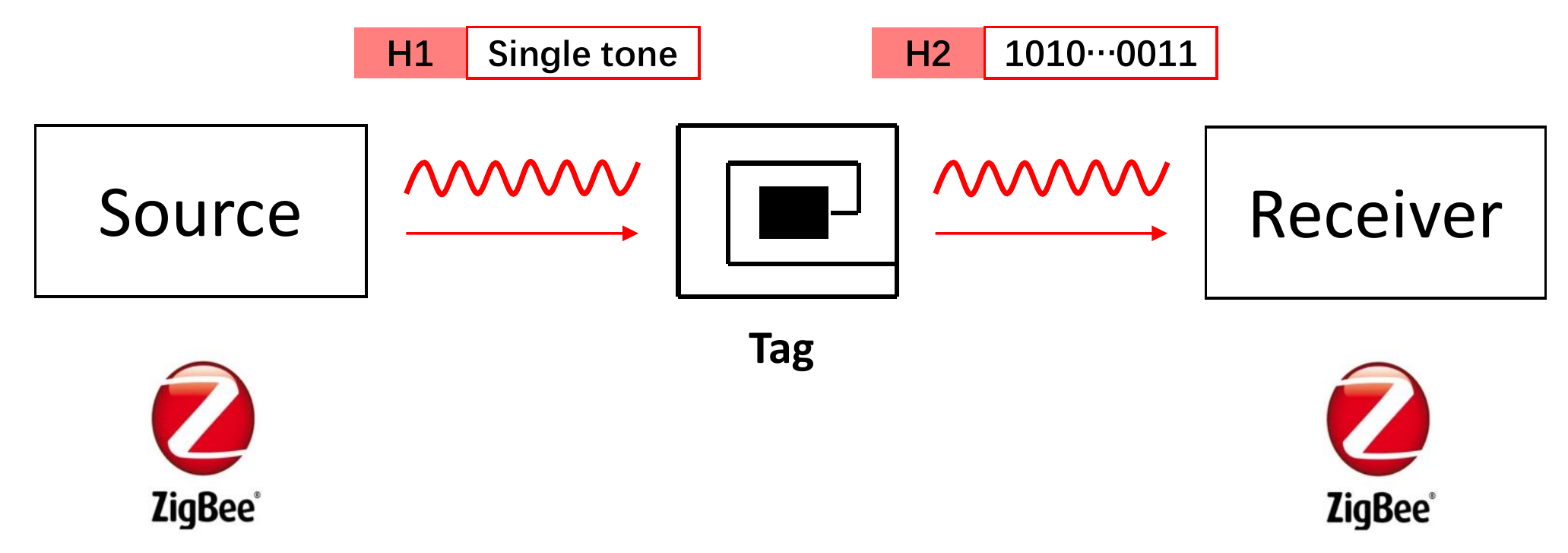}}
\caption{System structure. The commodity Zigbee transmitter transmits a Zigbee signal, which consists of a non-single-tone header and a single-tone payload. The backscatter tag uses the single-tone component as RF carrier and conducts FPS modulation on it to embed tag data. Then, the commodity Zigbee receiver will receive the backscattered signal and restore the tag data.}
\label{fig}
\end{figure}

\begin{figure}[tbp]
\centering
\centerline{\includegraphics[width=0.95\linewidth]{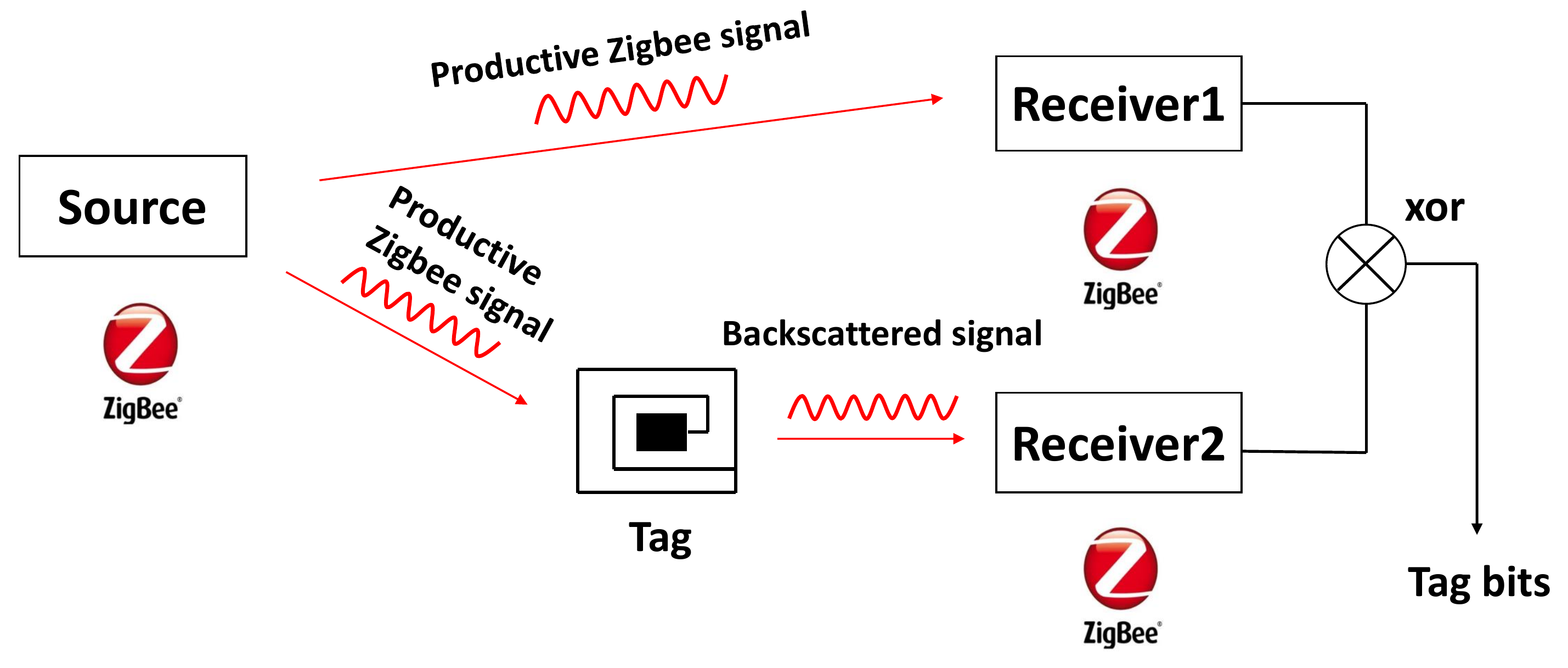}}
\caption{Freerider's System structure. The backscatter tag uses codeword translation to piggyback tag data. Two Zigbee receivers are set to recover the tag data. Specifically, the two receivers receive the carrier and the backscattered signal, respectively, and then restore the tag data by compare the codewords.}
\label{fig}
\end{figure}

In this paper, we applied FPS modulation to Zigbee backscatter system and successfully verified its feasibility. The structure of our system is displayed in Fig.1. The system consists of three components: a commodity Zigbee transmitter, a backscatter tag, and a commodity Zigbee receiver. The Zigbee transmitter transmits a Zigbee signal, which consists of a non-single-tone header and a single-tone payload. The tag uses the single-tone component as RF carrier and conducts FPS modulation on it. In addition, in order to prevent the carriers from interfering with the backscattered signals, we need to add an additional frequency shift $f_{shift}$ to move the backscattered signals to another channel.

Besides these, we creatively proposed to apply FPS modulation to Zigbee codeword translation. Freerider first proposed to use codeword translation for backscatter communication. The system structure is illustrated in Fig.2. Different from the traditional backscatter system showed in Fig.1, it has two receivers to recover the tag data and can use productive signals as carriers. It utilizes codeword translation to modulate tag data. As the core process of Zigbee codeword translation is to change the phase of the carrier, FPS can be used to improve the signal modulation on the tag.    

In conclusion, we made the following contributions in this paper:
\begin{itemize}
    \item We built a hardware prototype of a Zigbee backscatter system on an FPGA platform to realize FPS modulation. Through extensive experiments, we demonstrated that FPS modulated Zigbee transmissions have a bandwidth of 1.8 MHz, which is 2x lower than that of IPS modulated Zigbee transmissions.   
    \item We propose to apply FPS modulation to Zigbee codeword translation. The Section \Rmnum{3}-C explains our design. We have done some simulation experiments in MATLAB to prove its feasibility.
\end{itemize}

\section{Preliminaries}
\subsection{Zigbee Transceiver}
\textbf{Zigbee Transmitter.} The physical layer of Zigbee is IEEE 802.15.4. An active Zigbee transmitter uses Offset Quadrature Phase Shift Keying (OQPSK) modulation with a bitrate of 250 kbps and uses Direct Sequence Spread Spectrum (DSSS) to encode data. As illustrated in Fig.3(a), every four bits in Zigbee packets will be regarded as a symbol. The symbol will be spread into a 32-bit chip sequence, which will be modulated into an OQPSK signal to be transmitted.

\textbf{Zigbee Receiver.} The work process of a Zigbee receiver is shown in Fig.3(b). An Zigbee receiver uses Down Converter to extract the baseband signal, and then demodulate it according to the sign of phase shift between consecutive chip units. A positive phase shift will be demodulated to a bit '1' and a negative phase shift will be demodulated to a bit '0'. The demodulation rules are as follows:
\begin{equation}
\label{eq1}
result=\left\{
\begin{aligned}
1 & , &  &s(n)*s^*(n-1)\in [0,\pi] & \\
0 & , &  &s(n)*s^*(n-1)\in [-\pi,0]& 
\end{aligned}
\right.
\end{equation}

At last, the resulting sequence will be correlated with the mapping table. The receiver will find the chip sequence which has the minimum Hamming distance with the input sequence, and recover the symbol corresponding to it.

\begin{figure}[tbp]
\centering
\subfigure[Work Process of Active Zigbee Transmitter]
{
    \begin{minipage}[b]{1\linewidth}
        \centering
        \includegraphics[width=0.95\linewidth]{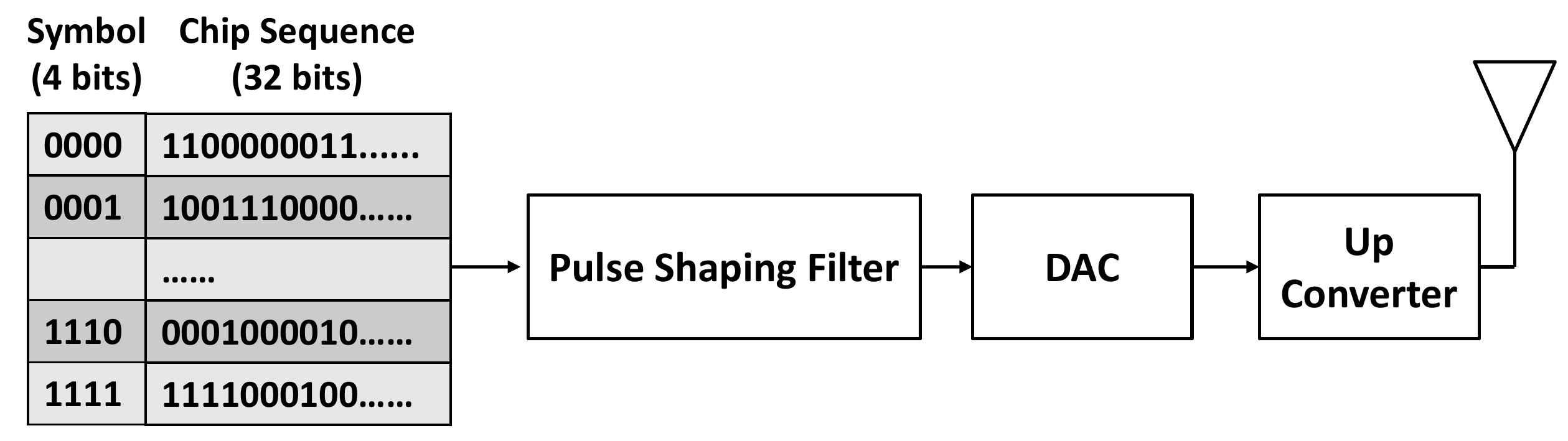}
    \end{minipage}
}

\subfigure[Work Process of Active Zigbee Receiver]
{
 	\begin{minipage}[b]{1\linewidth}
        \centering
        \includegraphics[width=0.95\linewidth]{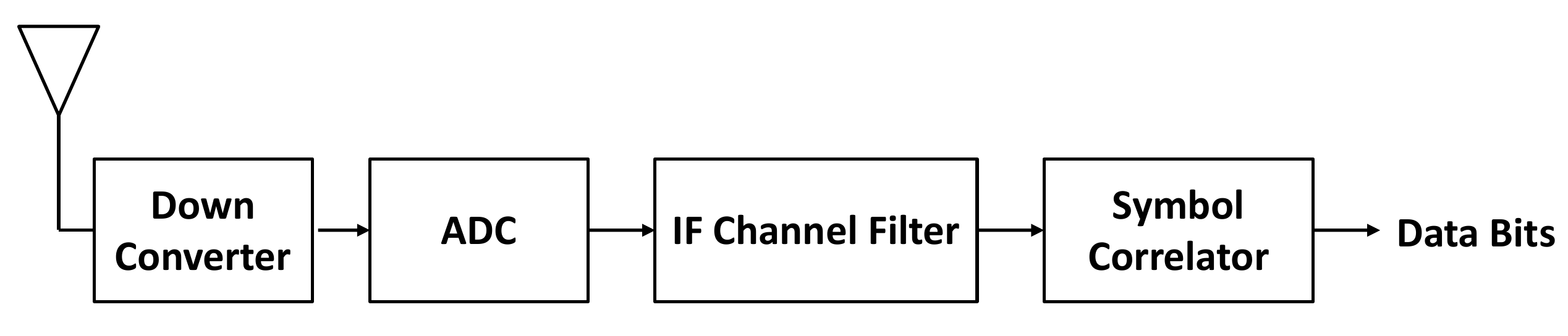}
    \end{minipage}
}
\caption{Work Process of Zigbee transceiver.}
\end{figure}

\subsection{Backscatter Communication}
A conventional backscatter system is illustrated in Fig.1. The system consists of three components: a carrier provider, a backscatter tag, and a backscatter receiver. The carrier provider transmits ambient signals (e.g., Wi-Fi, ZigBee, BLE, LTE, etc.) to the tag. The expression of the ambient signals is shown in Eq.(2). $A_{c}$, $f_{c}$, and $\phi_{c}$ are the amplitude, phase, and frequency of the ambient signals. The tag uses the ambient signals as carriers and mixes baseband signals with them to embed tag data. Specifically, when the carriers passing across the backscatter circuit, the tag will modify the reflection coefficient ($\Gamma$(t)). The expression of $\Gamma$(t) is shown in Eq.(3). $Z_{L}$ and $Z_{a}$ are the impedance of load and antenna, respectively. The backscattered signal B(t) is shown in Eq.(4). It can be considered as the product of carrier C(t) and reflection coefficient $\Gamma$(t).

\begin{align}
&C(t) = A_{c}e^{j(2\pi f_{c}t+\phi_{c})} \\
&\Gamma(t) = \frac{Z_{L}-Z_{a}^{*}}{Z_{L}+Z_{a}^{*}} \\
&B(t) = \Gamma(t)C(t)
\end{align}

\subsection{Zigbee Codeword Translation}
Conventional non-single-tone backscatter systems use codeword translation to piggyback tag data. The architecture of a Zigbee backscatter system is displayed in Fig.2. When the tag receives the Zigbee signal, it will use it as carrier and translate the carrier codeword into another codeword from the same codebook. Specifically, when the tag wants to piggyback a bit '1', it will reverse the bit in the carrier. On the contrary, if it wants to piggyback a bit '0', it will keep the bit unchanged. 
There are two receivers in the system to recover the tag data. One receiver receives carriers and the other receives backscattered signals. They restore the tag bits by comparing these two signals.

\subsection{Instantaneous-Phase Shift Modulation}
Instantaneous-Phase Shift (IPS) modulation is a popular modulation scheme to modify the phase of carriers and is widely used in advanced backscatter systems, such as interscatter. By toggling between square waves of different discrete phases, IPS can add phase shifts in instant. To be more specific, in interscatter's design, four square waves are used to modulate Zigbee-compatible signals. As displayed in Fig.4, there are four square waves that can be generated on the tag. All of these four square waves have a frequency of $f_{shift}$, which is used to shift the backscattered signals to another channel and can prevent carriers from interfering backscattered signals. The four square waves have different phases, which are 0, $\frac{\pi}{2}$, +$\pi$, +$\frac{3\pi}{2}$, respectively. The tag switches between these square waves to add phase shift to the carriers. For example, if we switch the square wave from phase 0 to phase $\frac{\pi}{2}$, we will add a phase shift of $\frac{\pi}{2}$ to the carrier.

\begin{figure}[tbp]
\centering
\centerline{\includegraphics[width=0.95\linewidth]{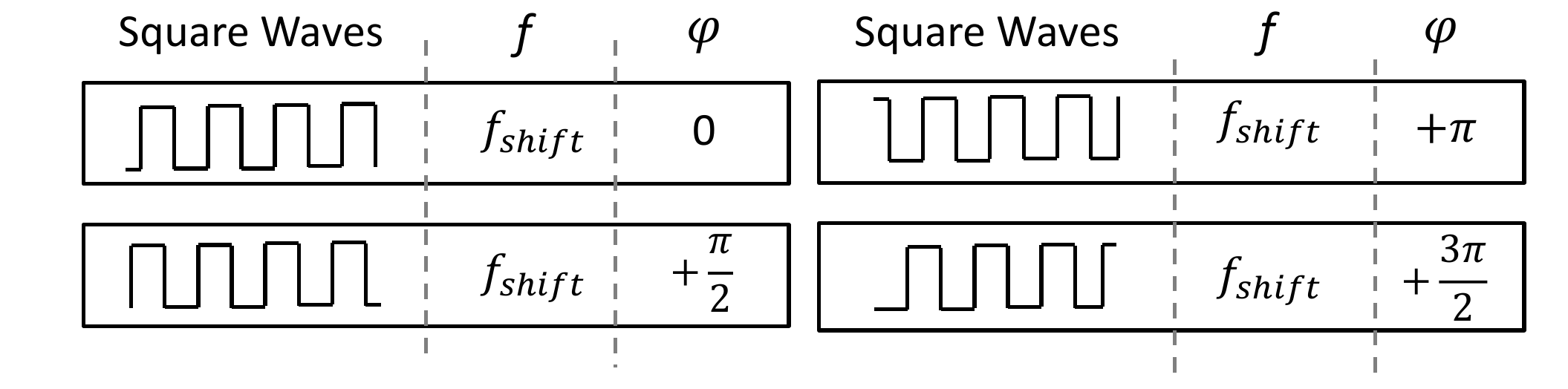}}
\caption{The four square waves that are used to IPS Modulation in interscatter design. These square waves have a frequency of $f_{shift}$, which is used to shift the backscattered signals to another channel. By toggling between these square waves, the tag can add phase shifts to the carreir.}
\label{fig}
\end{figure}

In the following, we will explain the process of IPS modulation by some expressions. As shown in Eq.(5), the fourier series of the square waves generated on the tag can be written as T(t). The expression of the backscattered signal (B(t)) is displayed in Eq.(6). It can be considered as the product of carrier C(t) and T(t). We only use the first term of T(t) to calculate B(t), because other terms will be eliminated by the receiver channel filter. The phase of the square wave is written as $\phi_{T}$ ($\phi_{T}\in \left \{ 0, \frac{\pi}{2}, +\pi, +\frac{3\pi}{2} \right \}$). By toggling between square waves of different phases, we can add phase shifts to the carrier.
The phase shift of IPS modulation is illustrated in Fig.5(a). In the first 0.5$\mu$s, we choose the square wave that have a phase of 0. In the second 0.5$\mu$s, we switch to the square wave whose phase is +$\frac{\pi}{2}$. Then, at the moment of switching, we add a phase shift of +$\frac{\pi}{2}$ to the carrier.
\begin{align}
T(t)& = A_{T} \sum_{n = 1,3,5...}^{} \frac{1}{n} e^{j(2\pi nf_{shift}t + \phi_{T})} \\
B(t)& = C(t)T(t) = A_{c}e^{j(2\pi f_{c}t+\phi_{c})}A_{T}e^{j(2\pi f_{shift}t + \phi_{T})}\\ \nonumber
& = A_{c}A_{T}e^{j(2\pi (f_{c}+f_{shift})t+(\phi_{c}+\phi_{T}))}
\end{align}

\begin{figure}[htbp]
\centering
\subfigure[IPS Modulation.]
{
    \begin{minipage}[b]{0.46\linewidth}
        \centering
        \includegraphics[width=1\linewidth]{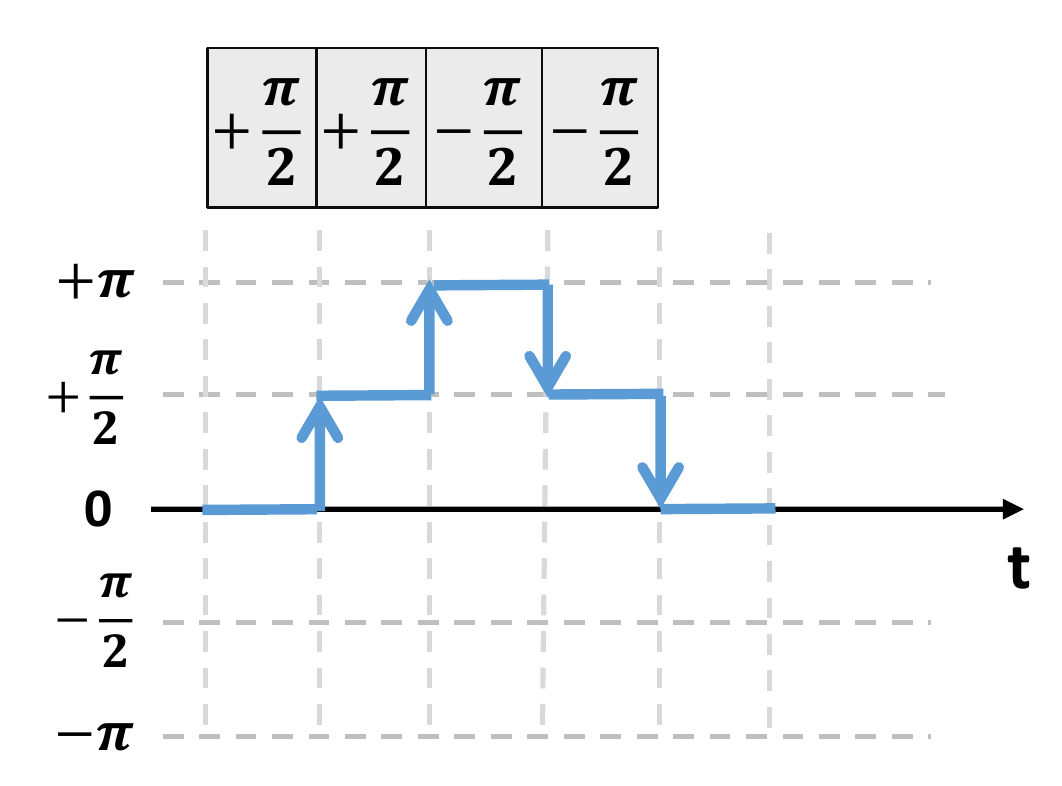}
    \end{minipage}
}
\subfigure[FPS Modulation.]
{
    \begin{minipage}[b]{0.46\linewidth}
        \centering
        \includegraphics[width=1\linewidth]{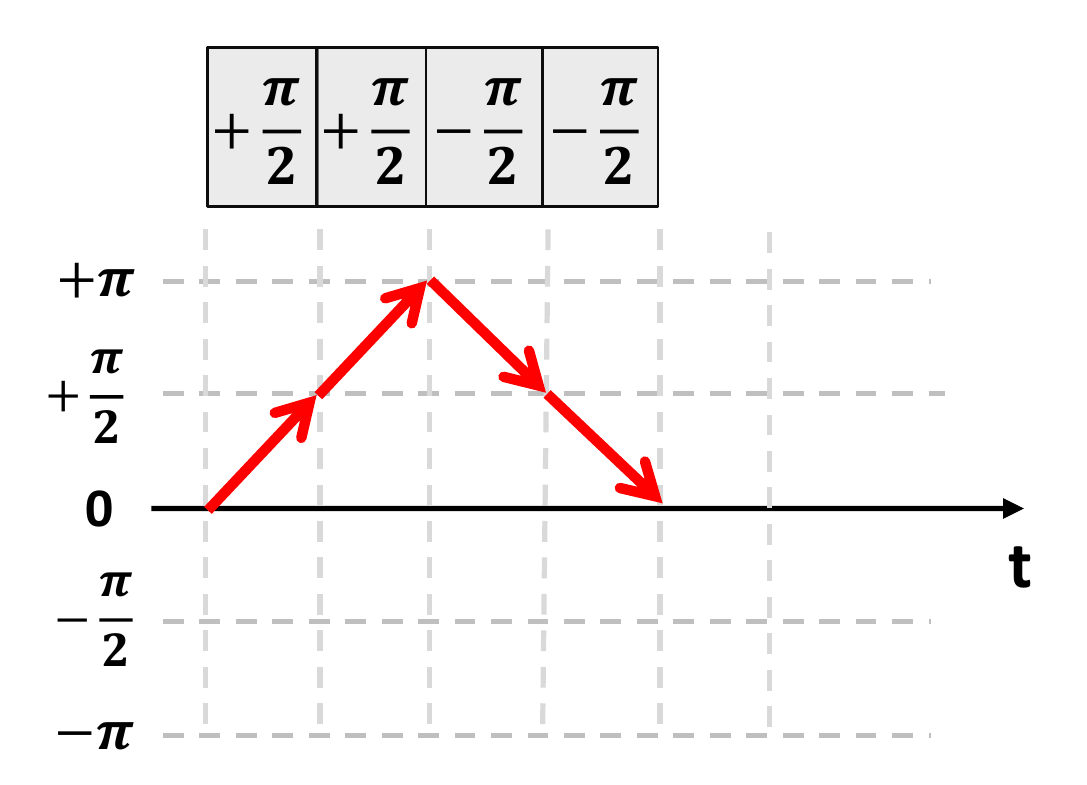}
    \end{minipage}
}
\caption{Phase shift of IPS Modulation and FPS Modulation.}
\end{figure}

\section{System Design}
In this paper, we propose FPS modulation, which is a fine-grained modulation and can modulate continuous phase shifts. In this section, we first introduce the design of FPS modulation. Then, we apply FPS modulation to a Zigbee backscatter system which uses single-tone Zigbee signals as carriers. At last, we propose to apply FPS modulation to Zigbee codeword translation.

\subsection{FPS design}
The core idea of FPS modulation is to add phase shift to carriers by modifying the frequency. FPS modulation produces continuous phase shifts and can suppress spectrum sidelobe. Similar to IPS modulation, FPS modulation piggyback tag data by switching between square waves. However, the square waves in FPS modulation have various frequencies($f_{shift} \pm f_{FP}$). The expression of FPS modulation is shown in Eq.(7). T(t) is the fourier series of the square waves generated on the tag. The expression of the backscattered signal B(t) is displayed in Eq.(8). It can be considered as the product of carrier C(t) and T(t). We only use the first term of T(t) to calculate B(t), because other terms will be eliminated by the receiver channel filter.

\begin{align}
T(t)& = A_{T} \sum_{n = 1,3,5...}^{} \frac{1}{n} e^{j(2\pi n(f_{shift} + f_{FP})t + \phi_{T})} \\
B(t)& = C(t)T(t)\\ \nonumber
&= A_{c}e^{j(2\pi f_{c}t+\phi_{c})}A_{T}e^{j(2\pi(f_{shift}+f_{FP})t + \phi_{T})}\\ \nonumber
& = A_{c}A_{T}e^{j(2\pi (f_{c}+f_{shift}+f_{FP})t+(\phi_{c}+\phi_{T}))}
\end{align}

As we can see from Eq.(7), the square waves generated on the tag have a frequency of $f_{shift}+f_{FP}$. We explain the meaning of $f_{shift}$ and $f_{FP}$ as follows: 1)\textbf{$f_{shift}$.} It is used to shift the backscattered signals to another channel, which can prevent carriers from interfering backscattered signals. 2)\textbf{$f_{FP}$.} It is set to add phase shift. Specifically, if we want to add a phase shift of $\Delta\phi$ within the transition time $\Delta T$, we should set $f_{FP}$ to $\frac{\Delta \phi }{2\pi \Delta T}$, and keep it for a time of $\Delta T$.

\subsection{Apply FPS to Zigbee Backscatter}
\textbf{Modulation on the tag.} As Zigbee uses Offset Quadrature Phase Shift Keying (OQPSK) modulation, we can utilize FPS modulation to achieve Zigbee backscatter. The architecture of our system is displayed in Fig.1. We use Zigbee single-tone signals as carriers and realize FPS modulation on the tag to transmit tag data to the Zigbee receiver. Specifically, FPS is achieved by toggling between square waves. As illustrated in Fig.6, there are eight square waves that can be generated on the tag. These square waves have two frequencies($f_{shift} \pm f_{FP}$) and four phases($0, \frac{\pi}{2}, +\pi, +\frac{3\pi}{2} $). As introduced before, the $f_{FP}$ can be written as $\frac{\Delta \phi }{2\pi \Delta T}$. Zigbee signals have a bandwidth of 2MHz and have a phase shift of $\pm\frac{\pi}{2}$ in every chip (0.5$\mu s$). Thus, $\Delta \phi = \frac{\pi}{2}$, $\Delta T = 0.5\mu s$, and $f_{FP} = 500KHz$.
$\phi_{T}$ is the initial phase of the square wave and is used to guaranteed phase continuity when switching square waves. For example, if we want to realize a phase shift($\frac{\pi}{2}\rightarrow\pi$), we should choose the square wave \#1. The phase shifts of FPS signals are shown in Fig.5(b). The phase shift is continuous, which can suppress spectrum sidelobe.

\textbf{Single-tone generation.} In our system, we used Zigbee single-tone signals as carriers. The process of generating Zigbee signals is shown in Fig.3(a). Zigbee uses OQPSK modulation. Thus, if we want to generate Zigbee single-tone signal, we should set the branch I to "101010..."($I(t) = cos(2\pi ft)$), and set the branch Q to "101010..."($Q(t) = sin(2\pi ft)$). The Zigbee signal can be written as $S(t) = I(t)+Q(t) = cos(2\pi ft)+sin(2\pi ft) = e^{j2\pi ft}$, which is a single-tone signal. However, Zigbee adheres to IEEE 802.15.4. It uses DSSS to encode data and spreads every four bits into one of the 32-bit chip sequences in the mapping table. Unfortunately, none of these chip sequence satisfies our needs. To generate Zigbee single-tone signals, we need to find a Zigbee radio that does not adhere to IEEE 802.15.4 strictly. We chose ATMEGA256RF2 radio as the carrier transmitter which is a Zigbee device that can generate Zigbee single-tone signal. Specifically, it supports various spreading factors(1, 2, 4, 8). When the spreading factor is set to 1, one bit is spread into one chip unit. The radio modulate Zigbee signal by frequency shift. A bit “0” represents a negative frequency deviation(-500 kHz), while a bit "1" represents a positive frequency deviation(+500 kHz). Thus, if we set the spreading factor to 1 and set the Zigbee packet to constant "0", we will get a Zigbee single-tone signal which has a frequency of $f_{c}-500k$.

\begin{figure}[tbp]
\centering
\centerline{\includegraphics[width=0.95\linewidth]{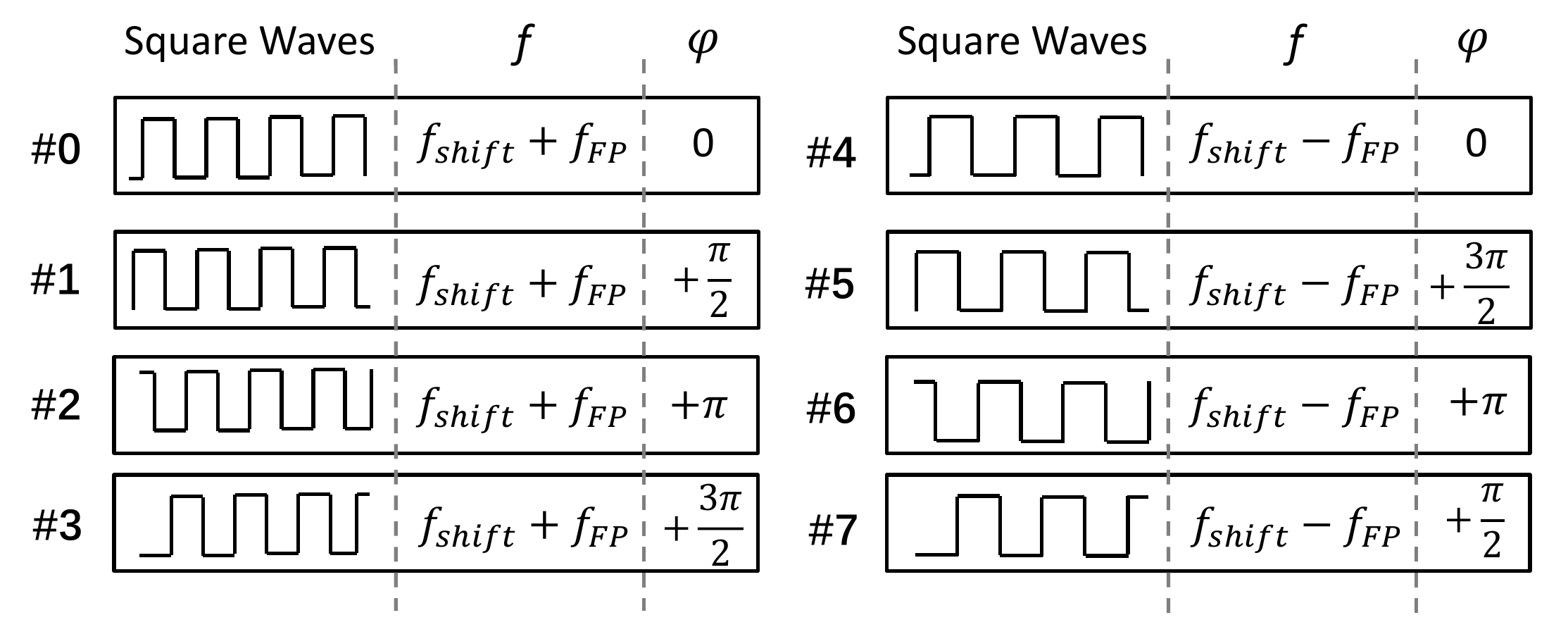}}
\caption{The eight square waves that are used in FPS Modulation. These square waves have two frequencies($f_{shift} \pm f_{FP}$) and four phases($0, \frac{\pi}{2}, +\pi, +\frac{3\pi}{2} $). The frequency $f_{shift}$ is used to shift the backscattered signals to another channel. The frequency $\pm f_{FP}$ is used to add phase shifts of $\pm \frac{\pi}{2}$. }
\label{fig}
\end{figure}

\subsection{Apply FPS to Zigbee Codeword Translation}
Freerider proposed to utilize codeword translation to achieve backscatter communication. The architecture of the backscatter system is displayed in Fig.2. The tag in the system piggybacks data by translating codeword. Specifically, when the tag wants to piggyback a bit '1', it will reverse the bit in the carrier. On the contrary, if it wants to piggyback a bit '0', it will keep the bit unchanged. In Zigbee signals, a bit '1' corresponding to a phase shift of $+\frac{\pi}{2}$ in 0.5$\mu$s, while a bit '0' corresponding to a phase shift of $-\frac{\pi}{2}$. Thus, if we want to reverse the bit in the carrier, we should add a phase shift of $+\pi$ to it in 0.5$\mu$s. According to these rules, we design a modulation scheme that can apply FPS modulation to Zigbee codeword translation, and construct a backscatter system similar to Freerider. Same as traditional FPS modulation, we toggle between square waves on the tag to modulate the backscattered signal. All of the square waves are illustrated in Fig.7. These square waves have two frequencies($f_{shift}+f_{FP}, f_{shift}$) and two phases($0, +\pi$). $f_{FP}$ can be written as $\frac{\Delta \phi }{2\pi \Delta T}$. As we need to add a phase shift of $+\pi$ in 0.5$\mu$s when piggybacking a bit '1', $\Delta \phi = +\pi$, $\Delta T = 0.5\mu s$, and $f_{FP} = 1MHz$. When piggybacking a bit '1' on the tag, it should switch to the square waves that have a frequency of $f_{shift}+f_{FP}$. Otherwise, it should switch to the square waves that have a frequency of $f_{shift}$, which do not influence the phase of the carrier.

\begin{figure}[tbp]
\centering
\centerline{\includegraphics[width=0.95\linewidth]{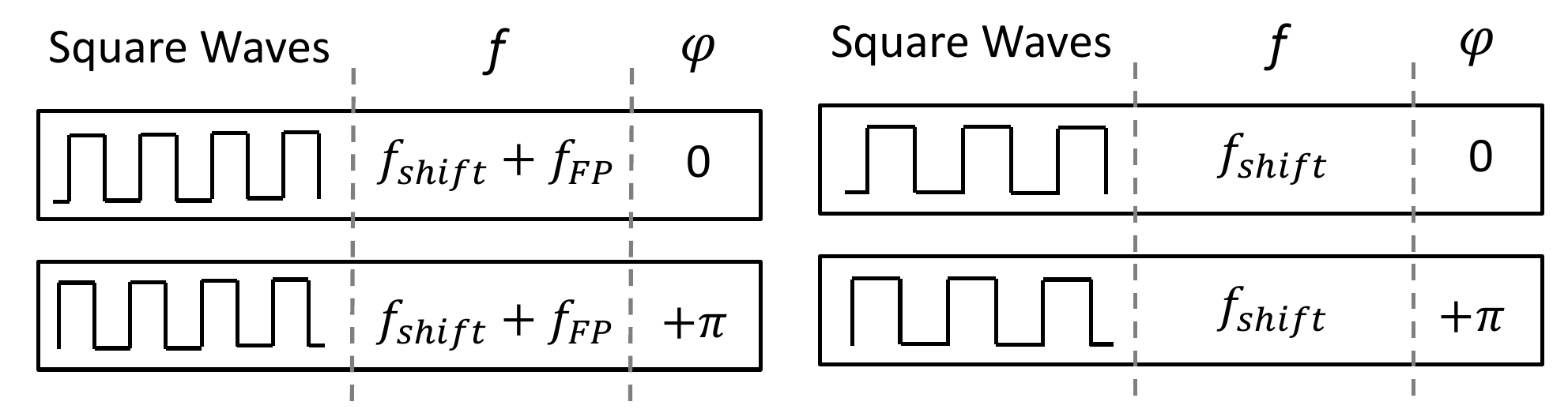}}
\caption{The four square waves that are used in Zigbee codeword translation. These square waves have two frequencies($f_{shift}+f_{FP}, f_{shift}$) and two phases($0, +\pi$). The frequency $f_{shift}$ is used to shift the backscattered signals to another channel. The frequency $f_{FP}$ is used to add phase shifts of $+\pi$. }
\label{fig}
\end{figure}

\section{Implementation}
\textbf{Zigbee transceiver.} We use an ATMEGA256RF2 radio as carrier transmitter and an off-the-shelf TI CC2650
as ZigBee receiver. The Zigbee transmitter sets the spreading factor to be 1 and sets the payload with constant “0". Then, it will generate a negative single tone signal, whose frequency is -500 kHz from the central frequency. Our transmitter works on Zigbee channel 12, whose central frequency is 2410MHz. Our backscatter tag can modulate IEEE 802.15.4 Zigbee packets, so the backscattered signal can be successfully demodulated by a commodity Zigbee receiver.

\textbf{Backscatter tag.} We prototype the backscatter tag by a RF front-end circuit and an FPGA. The RF front-end circuit consists of three components: envelope detector, a comparator, and an RF switch. The envelope detector is AD8313. It gets the envelope and transmit the output to the comparator. The comparator is used to  eliminate noise and sets a threshold for downlink instructions decoding. The RF switch is ADG902. It switches different impedance according to the FPGA's output to generate backscattered signals. The FPGA in our system is XILINX ZYNQ 7000. It processes the baseband signals and modulate the backscattered signals on ZigBee channel 14(2420MHz). Thus, the $f_{shift}$ should be set to 10MHz.

\section{Evaluation}
\subsection{Single-tone Generation}
In Section \Rmnum{3}-B, we have introduced our method to generate Zigbee single-tone. In this section, we uses ATMEGA256RF2 radio to generate Zigbee single-tone. We connect the antenna of the radio to a PXIe-5663 RF Vector Signal Analyzer (VSA). The VSA is used to observe spectrum, and the experimental results are illustrated in Fig.8. The results show that if we set PSDU to constant "0"s, we will generate a Zigbee single-tone signal whose frequency is -500 kHz from the central frequency. On the contrary, if we set PSDU to constant "1"s, the frequency of the signal is +500 kHz from the central frequency.

\begin{figure}[htbp]
\centering
\subfigure[PSDU = "000..."]
{
    \begin{minipage}[b]{0.46\linewidth}
        \centering
        \includegraphics[width=1\linewidth]{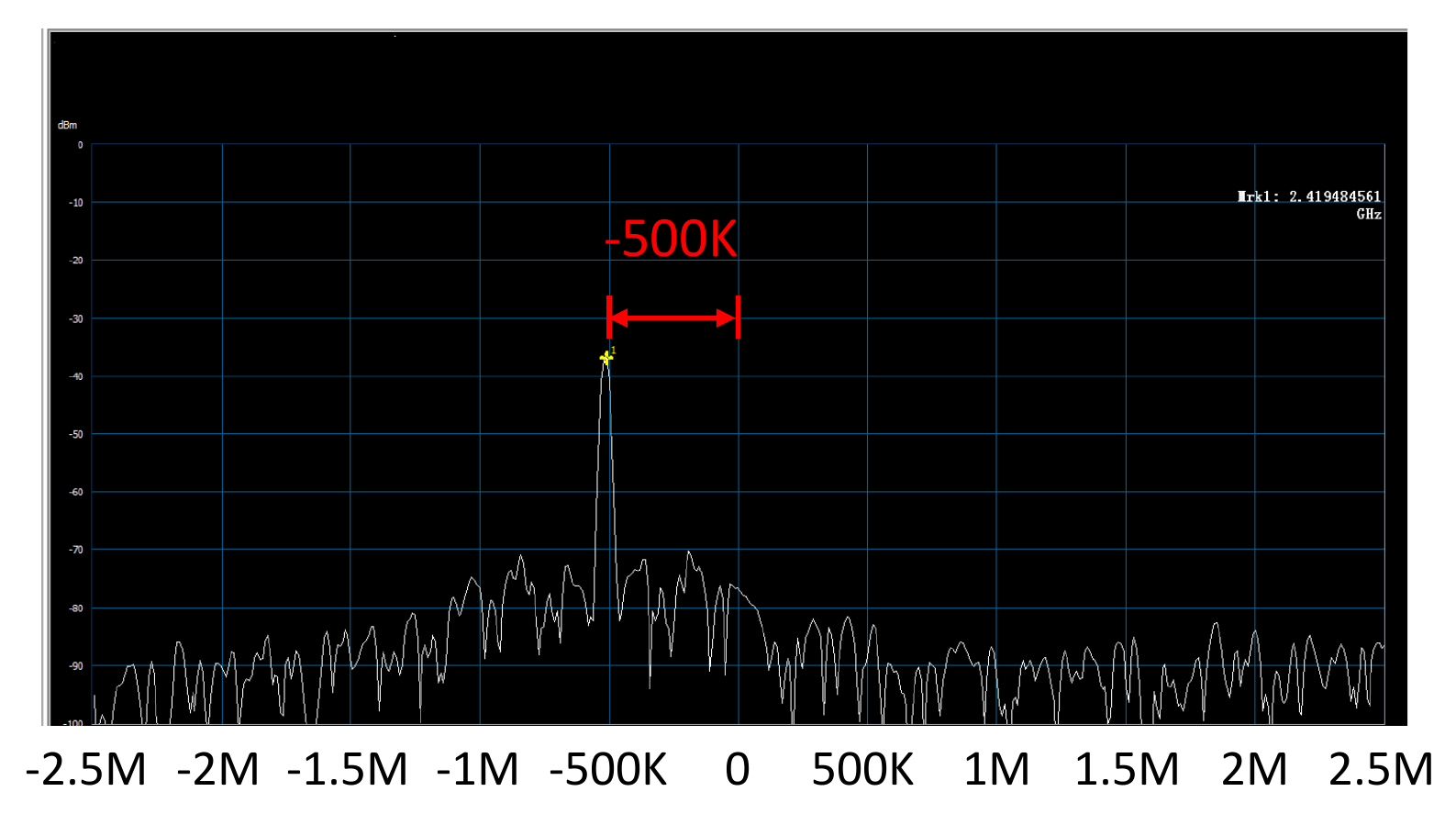}
    \end{minipage}
}
\subfigure[PSDU = "111..."]
{
    \begin{minipage}[b]{0.46\linewidth}
        \centering
        \includegraphics[width=1\linewidth]{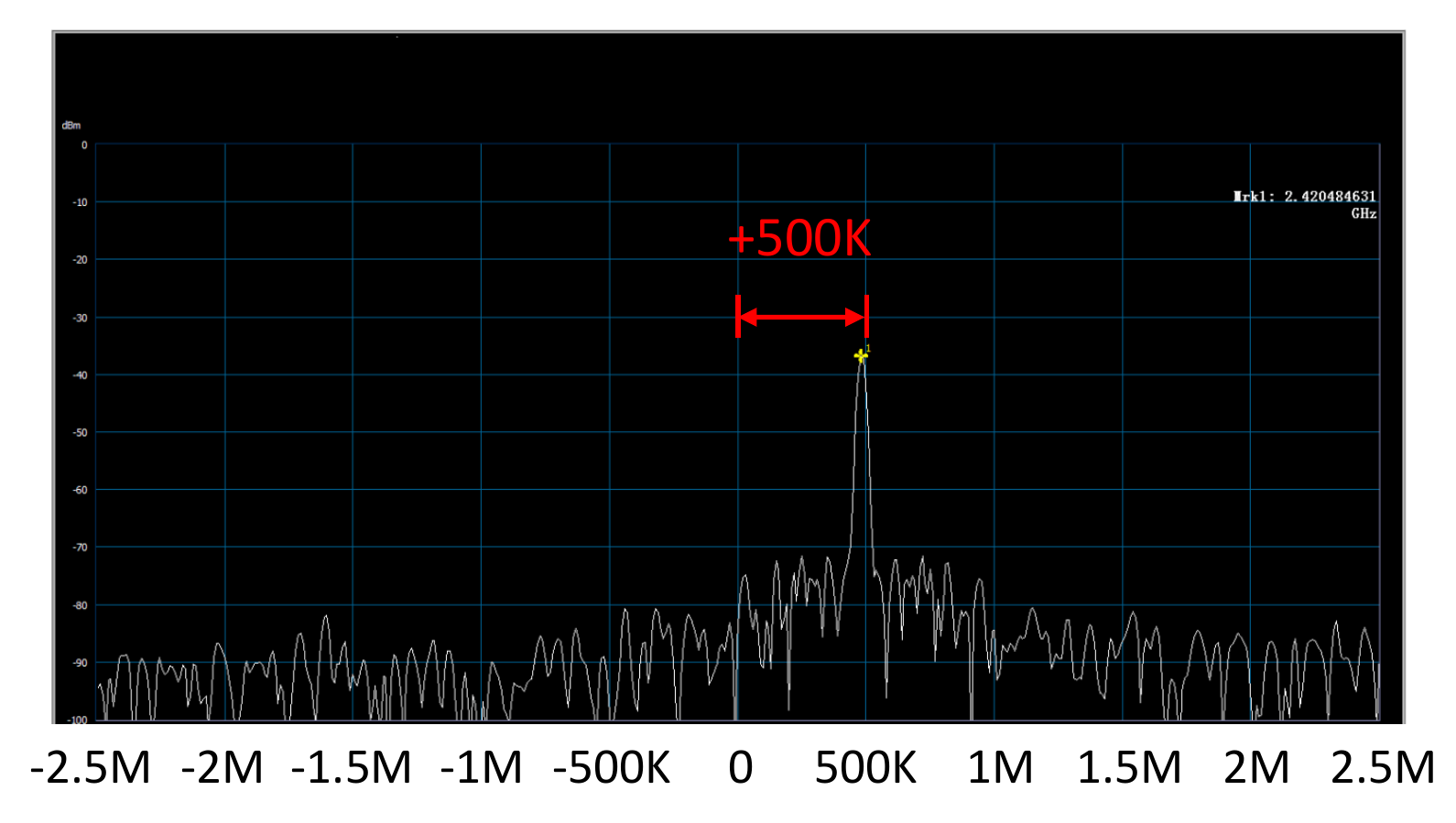}
    \end{minipage}
}
\caption{The spectrum of single-tone carriers.}
\end{figure}

\subsection{Spectrum Efficiency Comparison}
We evaluate the bandwidth of IPS modulation and FPS modulation, respectively. A commodity Zigbee radio (ATMEGA256RF2) transmits IEEE 802.15.4 packets continuously. Our tag takes single-tone carriers to transmit packets with IPS and FPS. The output of the tag is connected to a PXIe-5663 RF Vector Signal Analyzer (VSA), which is used to measure the bandwidth. The experimental results are shown in Fig.9. We measure the occupied bandwidth containing over 90\% RF energy. The IPS modulated signals have a bandwidth of 3.2MHz, while the FPS modulated signals have a bandwidth of 1.8MHz. The results demonstrated that the bandwidth of FPS is about 2x better than IPS, which means that FPS has a better spectrum efficiency. 

\begin{figure}[htbp]
\centering
\subfigure[FPS modulation]
{
    \begin{minipage}[b]{0.46\linewidth}
        \centering
        \includegraphics[width=1\linewidth]{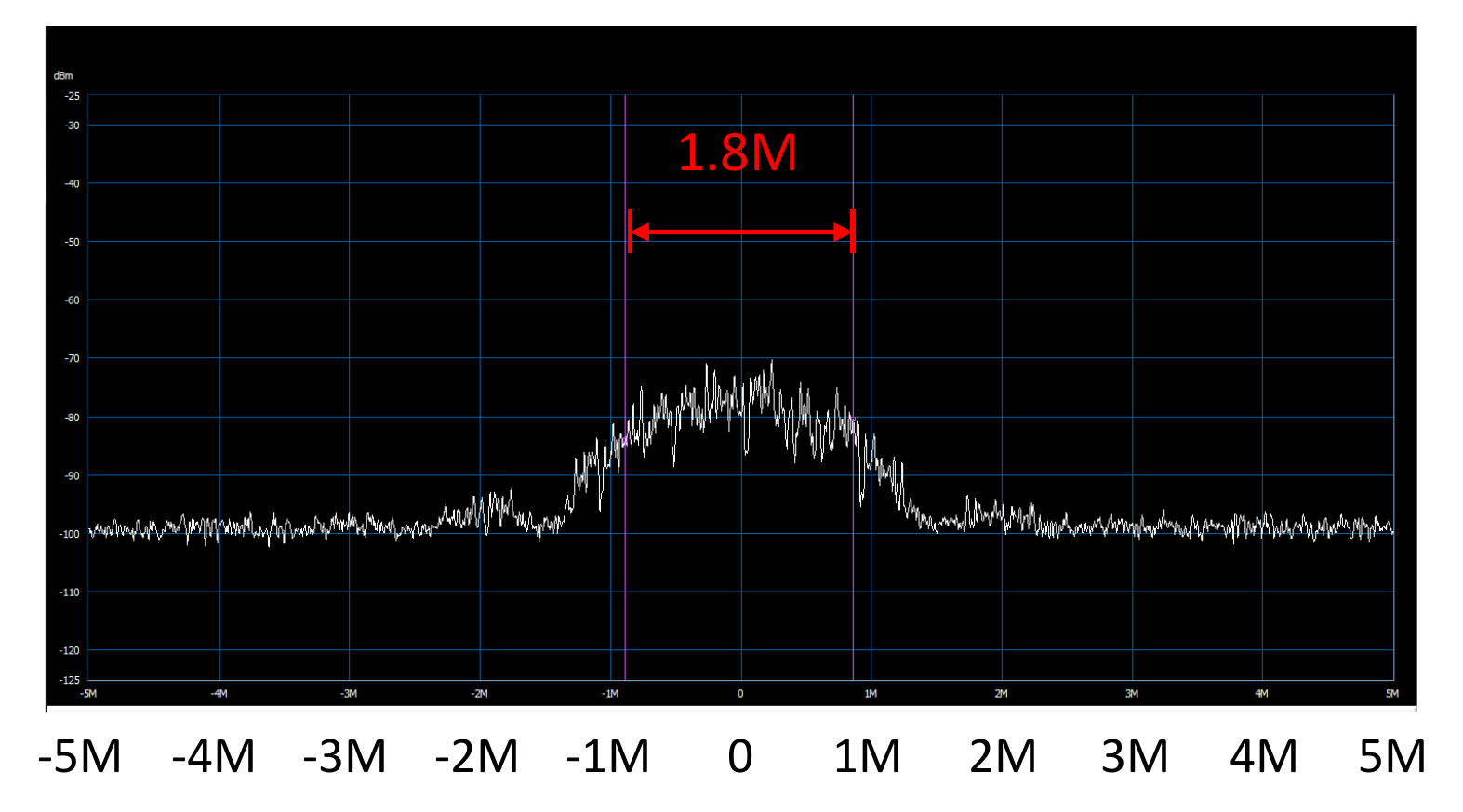}
    \end{minipage}
}
\subfigure[IPS modulation]
{
    \begin{minipage}[b]{0.46\linewidth}
        \centering
        \includegraphics[width=1\linewidth]{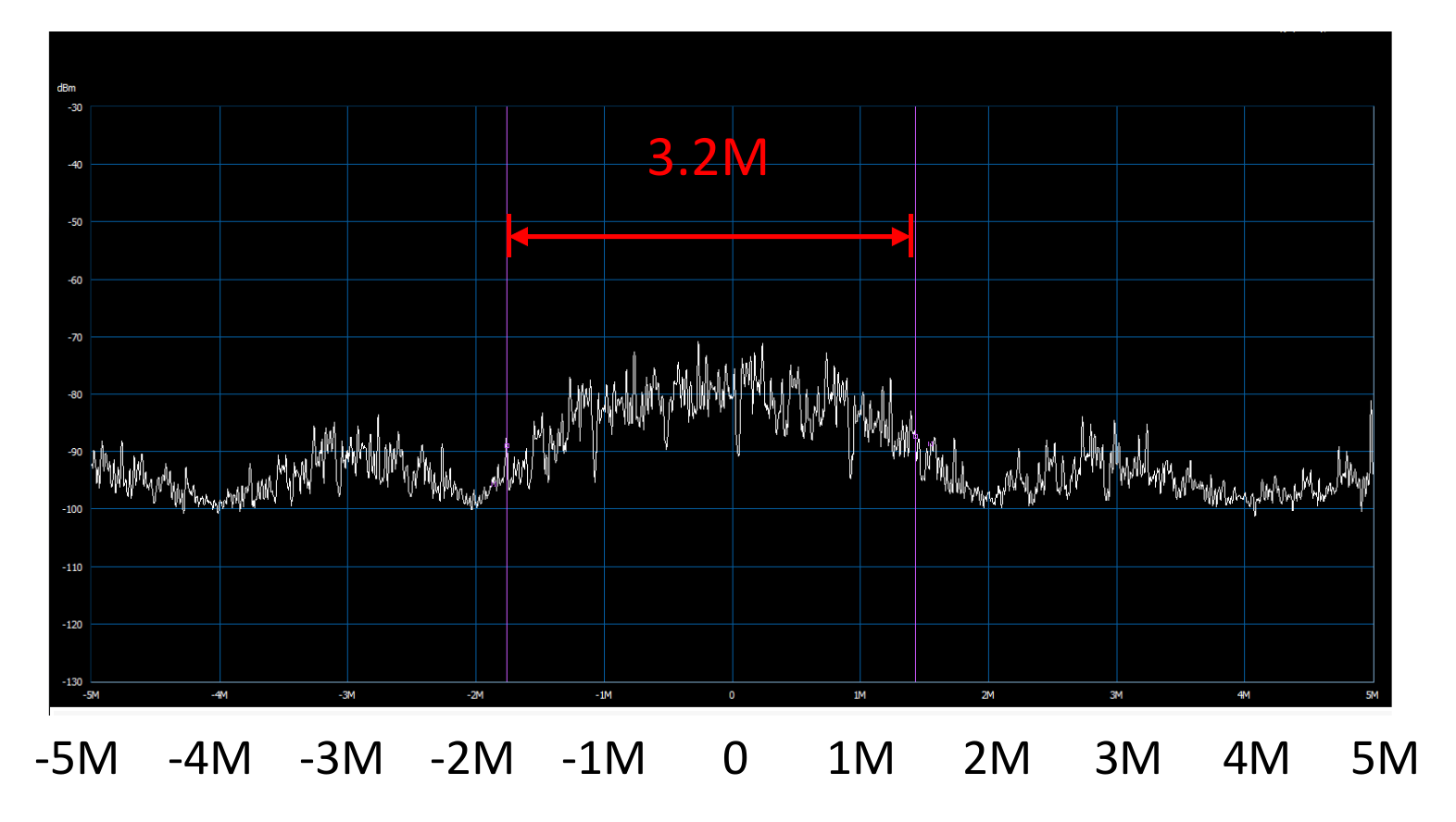}
    \end{minipage}
}
\caption{The bandwidth of IPS modulation and FPS modulation.}
\end{figure}

\end{document}